\documentclass[12pt]{article}
\usepackage[T1]{fontenc}
\usepackage{geometry}
\usepackage{graphics}

\makeatletter

\providecommand{\LyX}{L\kern-.1667em\lower.25em\hbox{Y}\kern-.125emX\@}

\makeatother
\begin{document}

\title{Processes \( e^{+}e^{-}\rightarrow c\bar{c}c\bar{c} \) and \( e^{+}e^{-}\rightarrow J/\psi +gg \)
at \( \sqrt{s}=10.59 \) GeV.}

\author{
A.V.Berezhnoy
\thanks{Scobeltsyn Institute for Nuclear Physics of Moscow State University,
Moscow, Russia.
}, 
A.K.Likhoded\thanks{Institute for High Energy Physics, Protvino, Russia.}}

\date{{}}

\maketitle
\begin{abstract}
BELLE Collaboration data for \( J/\psi  \) inclusive production in
the processes \( e^{+}e^{-}\rightarrow J/\psi +gg \) and \( e^{+}e^{-}\rightarrow J/\psi +c\bar{c} \)
have been discussed. These data have been compared with the predictions
of two pQCD methods: the former method use the information about the
\( J/\psi  \) wave function, the latter one do not use such information
and is based on the quark-hadron duality hypothesis. Both these calculation
methods lead to essential discrepancy between the theoretic predictions
and the experimental data. The production cross section dependence
on effective gluon mass has been investigated for the process \( e^{+}e^{-}\rightarrow J/\psi +gg \).
The production cross section of doubly charmed baryons \( \Xi ^{*}_{cc} \)
has been estimated.
\end{abstract}

\section{Introduction.}

Perturbative QCD allows us to describe the hard part of hadron production
process. The most essential success is achieved in the description
of the heavy meson production and the heavy quarkonium production
at large transverse momentum, where heavy quarks is produced in the
hard process followed by the soft hadronization process. The hadronization
is described by common coefficient which is slightly depends on transverse
momentum \( p_{T} \). It is not unlikely, that the final state interaction
can strongly brake the factorization at small transverse momenta and
small energies. The existence of charge asymmetry in the charm particle
production confirm this supposition. Thus, we expect the deviation
from the common pQCD fragmentation mechanism at small energies and
small \( p_{T} \). So, the recent BELLE Collaboration research \cite{Belle}
of pair production \( J/\psi +\eta _{c} \) and inclusive production
\( J/\psi +c\bar{c} \) in the \( e^{+}e^{-} \)-annihilation at \( \sqrt{s}=10.6 \)
GeV results in the production cross section values, which exceed the
pQCD predictions \cite{KLShev,Leibovich,2charmonium} by order of magnitude. The
experimental shapes of some cross section distribution for the process
\( e^{+}e^{-}\rightarrow J/\psi +gg \) crucially differ from the
pQCD predictions. 

It was shown in recent works \cite{Bodwin}, that the account of electromagnetic
\( J/\psi +J/\psi  \) and \( J/\psi +\eta _{c} \) pair production
increase almost twice the theoretical estimation of the cross section.
However, it does not remove the existing discrepancy between the pQCD
predictions and the experimental data for \( c\bar{c}c\bar{c} \)
production process. One more disagreement takes place for the prompt
\( J/\psi  \) production in the processes \( e^{+}e^{-}\rightarrow J/\psi +c\bar{c} \)
and \( e^{+}e^{-}\rightarrow J/\psi +gg \): the experimental value
\cite{Belle}

\begin{equation}
\label{ratio}
\sigma (J/\psi +c\bar{c})/\sigma (J/\psi +gg)=0.59_{-0.13}^{+0.15}\pm 0.12
\end{equation}

contradicts pQCD estimations\cite{Leibovich}\[
\sigma (J/\psi +c\bar{c})/\sigma (J/\psi +gg)\sim 0.1.\]

It is worth to mention, that the both cross section \( \sigma (J/\psi +c\bar{c}) \)
and \( \sigma (J/\psi +gg) \) are the same order of \( \alpha _{s} \)
and at \( \sqrt{s}=10.6 \) GeV slightly depend on the used models,
as well as on the chosen scale. 

In this article we shall discuss in detail the existing discrepancies
between the pQCD predictions and the experimental data.

\section{Process \protect\( e^{+}e^{-}\rightarrow J/\psi +c\bar{c}\protect \).}

The experimental values of the production cross section for the inclusive
process \begin{equation}
\label{Jpsicc}
e^{+}e^{-}\rightarrow J/\psi +c\bar{c},
\end{equation}
which are reconstructed from the \( e^{+}e^{-}\rightarrow J/\psi +D^{0}+X \)
and \( e^{+}e^{-}\rightarrow J/\psi +D^{*+}+X \) equal to \begin{equation}
\label{sigJpsicc}
\sigma (e^{+}e^{-}\rightarrow J/\psi c\bar{c})=1.1_{-0.30}^{+0.36}\pm 0.26\; \mbox {pb}\; \mbox {and}\; 0.74_{-0.24}^{+0.28}\pm 0.19\; \mbox {pb},
\end{equation}
 correspondingly\cite{Belle}.

\begin{figure}
{\centering \resizebox*{0.7\textwidth}{!}{\includegraphics{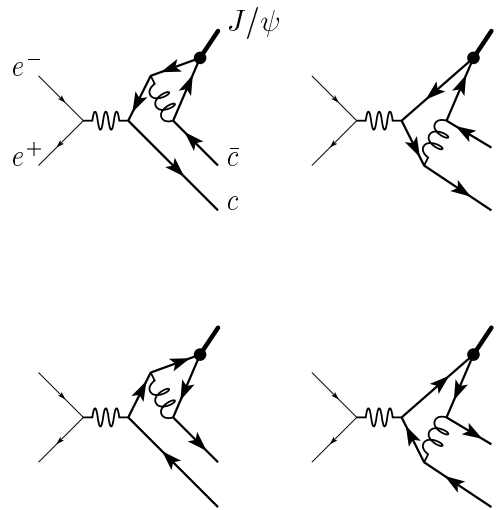}} \par}

\caption{Feynman diagrams for the process \protect\( e^{+}e^{-}\rightarrow J/\psi +c\bar{c}\protect \).
\label{ee}}
\end{figure}

In the frame work of pQCD the process (\ref{Jpsicc}) is described
by the diagrams of Fig.~\ref{ee}. The process amplitude can be represented
as the hard \( c\bar{c} \) pair production amplitude multiplied by
amplitudes of the soft fusion into \( c\bar{c} \)-quarkonium. Such
estimations were done independently by several research groups \cite{KLShev,Leibovich,Baek}.
These calculations are in a good accordance to each other and gives
the cross section value %
\footnote{The recent calculations \cite{Liu} for \( \alpha _{s}=0.26 \) results
in twice as much value of this cross section.
}: \begin{equation}
\label{sigJpsiccqcd}
\sigma (J/\psi +c\bar{c})\simeq 6\cdot 10^{-2}\; \mbox {pb},
\end{equation}
 which underestimates the experimental data (\ref{sigJpsicc}) by
the oder of magnitude. The cross section (\ref{sigJpsiccqcd}) increase
due to the increase of \( \alpha _{s} \) do not eliminate this discrepancy.
The factorization approach is not most likely reason of the disagreement.
It was shown in work \cite{KLShev}, that the factorization approach
predicts practically the same cross section values, as ones calculated
in the frame work of the method, which is based on the quark-hadron
duality hypothesis, and where no factorization, no information about
wave function are used. 

So, \( c\bar{c}c\bar{c} \)-production cross section values calculated
for the color singlet state of \( c\bar{c} \)-pair in the duality
interval \( 2m_{c}\leq M_{c\bar{c}}\leq 2M_{D^{*}}+\Delta M \) (\( \Delta M\simeq 0.5\div 1 \)
GeV) for \( \alpha _{s}=0.24 \) and \( m_{c}=1.4 \) GeV equal to
\cite{KLShev}:

\begin{equation}
\label{sig05}
\sigma _{c\bar{c}}(\Delta M=0.5)=0.13\; \mbox {pb}
\end{equation}

\begin{equation}
\label{sig1}
\sigma _{c\bar{c}}(\Delta M=1.0)=0.16\; \mbox {pb}
\end{equation}

It is worth to compare these results with the production cross section
of all \( S \)-wave quarkonium states:

\begin{equation}
\label{sigS}
\sigma (\Sigma \eta _{c,}\psi )=0.13\; \mbox {pb}
\end{equation}

It is clear from (\ref{sig05}, \ref{sig1}) and (\ref{sigS}) that,
the rough estimation of the cross section for the process (\ref{Jpsicc}),
based on duality hypothesis, as well as the more rigorous factorization
method lead to same cross section value. Nevertheless, as it was mentioned
in Introduction, this prediction underestimates BELLE experimental
data by order of magnitude.

It is necessary to note another important circumstance. The total
\( c\bar{c}c\bar{c} \) production cross section calculated by us
at \( \alpha _{s}=0.24 \), \( m_{c}=1.4 \) GeV and \( \sqrt{s}=10.6 \)
equals to \begin{equation}
\label{sig4c}
\sigma (c\bar{c}c\bar{c})=0.237\; \mbox {pb},
\end{equation}
which makes up \( W_{c\bar{c}}=2\times 10^{-4} \) of the total cross
section of the \( c\bar{c} \)-pair production at the same energy.
At \( Z^{0} \) boson peak, where the interaction energy is higher
by order of magnitude, the value of \( W_{c\bar{c}} \) is about \( 0.03 \).
Such behavior of the probability \( W_{c\bar{c}} \) to produce the
additional \( c\bar{c} \)-pair is described by pQCD calculation well
\cite{KLShev-2}. It should be noted that (\ref{sig4c}) is less than
experimental cross section value for the process \( J/\psi +c\bar{c} \).
The latter circumstance could indicate the strong suppression of the
four \( D \)-meson production. 

The cross section of the electromagnetic \( c\bar{c}c\bar{c} \) production
forms about 2.5\% of the total production cross section (\ref{sig4c})
: \[
\sigma ^{QED}(c\bar{c}c\bar{c})\simeq 6.6\cdot 10^{-3}\; \mbox {pb}\]

The theoretical predictions for the cross section of the electromagnetic
production of the \( S \)-wave states \cite{Bodwin}%
\footnote{
The cross section value of the electromagnetic production of \( S \)-wave
state estimated in the recent work \cite{Luchinsky} is nearly two
times less than predicted in \cite{Bodwin}.
} are several times greater than experimental value of \( c\bar{c}c\bar{c} \)
production. Perhaps, this fact indicates the underestimation of the
experimental value of \( \sigma ^{QED}(c\bar{c}c\bar{c}) \). Also,
it is not unlikely, that electromagnetic mechanism of the \( c\bar{c}c\bar{c} \)
production is essentially differ from QCD production mechanism.

\section{Process \protect\( e^{+}e^{-}\rightarrow J/\psi +gg\protect \).}

\begin{figure}
{\centering \resizebox*{0.7\textwidth}{!}{\includegraphics{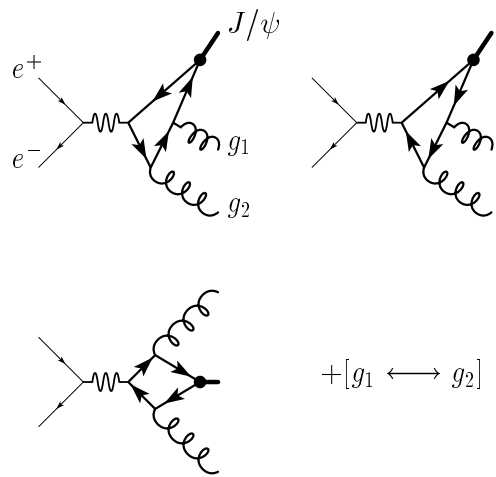}} \par}

\caption{Feynman diagrams for the process \protect\( e^{+}e^{-}\rightarrow J/\psi +gg\protect \).\label{jpsigg}}
\end{figure}

Let us consider one more process, which contributes to the inclusive
\( J/\psi  \) production (see diagrams in Fig.~\ref{jpsigg}):\[
e^{+}e^{-}\rightarrow J/\psi +gg.\]

To study this process in detail one need to read the papers\cite{All_jpsigg}.

The process cross section calculated in the frame work of color singlet
model at \( \sqrt{s}=10.6 \) GeV and \( \alpha _{s}=0.24 \) equals
to\begin{equation}
\label{sigJpsiggqcd}
\sigma (J/\psi +gg)=0.7\; \mbox {pb}.
\end{equation}

The estimation of \( J/\psi +c\bar{c} \) production in the frame
work of the same model leads to the following ratio between the cross
sections: \begin{equation}
\label{theor-ratio}
\sigma (J/\psi +c\bar{c})/\sigma (J/\psi +gg)=6\cdot 10^{-2}/0.7\approx 0.1
\end{equation}
It is necessary to compare this ratio value with the experimental
one (\ref{ratio}). The prediction uncertainty of ratio (\ref{theor-ratio})
is smaller than the cross section uncertainties themselves, because
(\ref{theor-ratio}) does not include the nonrelativistic amplitude
of the soft process \( c\bar{c}\rightarrow J/\psi  \). Moreover,
the both processes \( e^{+}e^{-}\rightarrow J/\psi +c\bar{c} \) and
\( e^{+}e^{-}\rightarrow J/\psi +gg \) are characterized by approximately
the same virtualities, that is why the ratio (\ref{theor-ratio})
does not depend on factorization scale chosen to calculate \( \alpha _{s} \). 

\begin{figure}
\( \displaystyle \frac{d\sigma _{J/\psi +gg}}{dm_{gg}} \), pb/GeV

{\centering \resizebox*{0.7\textwidth}{!}{\includegraphics{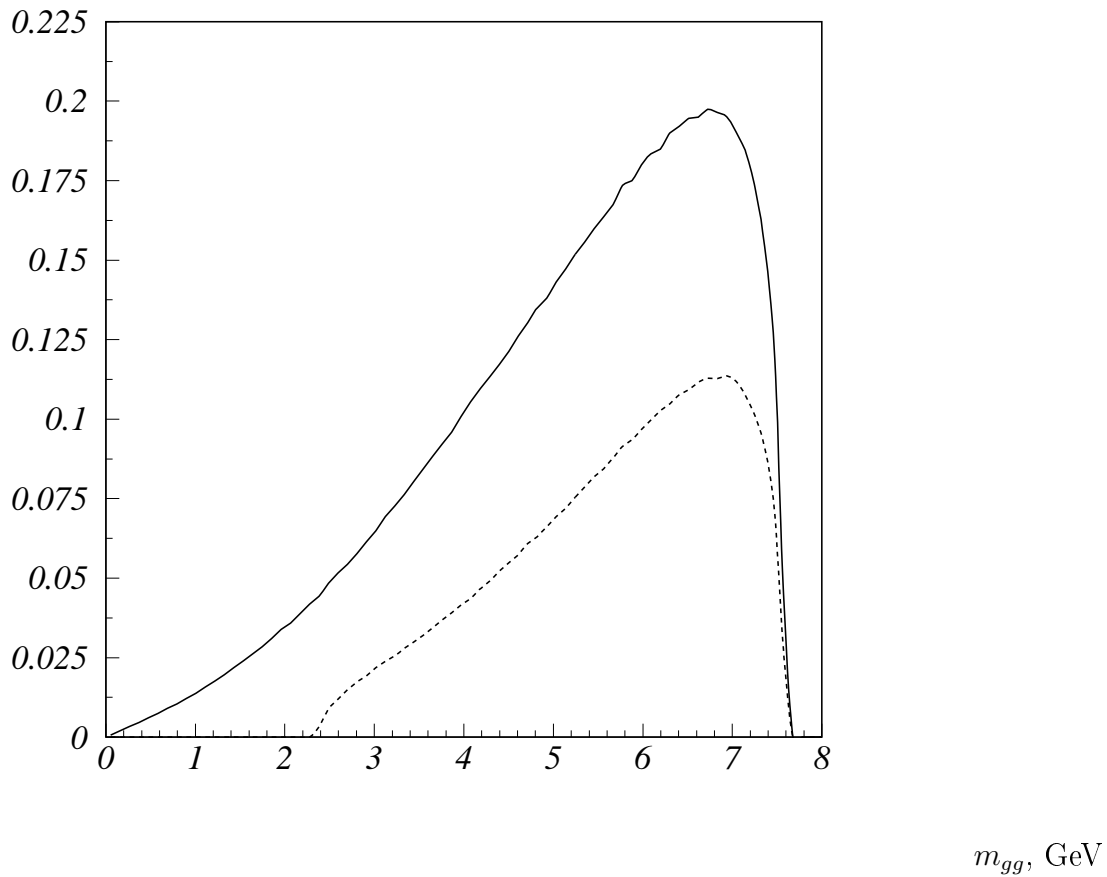}} \par}

\hfill \( m_{gg} \), GeV

\caption{The cross section distributions on \protect\( m_{gg}\protect \)
for \protect\( J/\psi +gg\protect \)-production in \protect\( e^{+}e^{-}\protect \)-annihilation
for massless gluons (\protect\( m_{g}=0\protect \), solid curve),
as well as for massive ones (\protect\( m_{g}=1.18\protect \) GeV,
dashed curve).\label{mgg}}
\end{figure}

It is worth to point out, that some difficulties exist for the prompt
using of the pQCD to predict the \( J/\psi +gg \) production. It
is well known that the essential discrepancy exits between the LO
pQCD predictions and the experimental data at small digluon masses
\( m_{gg} \) for decays \( J/\psi \rightarrow \gamma +gg \) and
\( \Upsilon \rightarrow \gamma +gg \). It was shown in \cite{leibovich-2}
that taking into account next order in the calculations strongly modify
the spectrum at small digluon masses. The analogous effect can be
achieved by using the effective nonzero mass of gluon. To estimate
the gluonic mass correction for the \( J/\psi +gg \) production cross
section we follow the work \cite{Field} and choose the gluonic mass
to be 1.18 GeV (as it was shown in\cite{Field}, this mass value is
needed to describe the photonic spectrum in the decay \( \Upsilon \rightarrow \gamma +X \)).
In our case the nonzero gluon mass shifts the \( m_{gg} \) spectrum
to the region of large \( m_{gg} \) values and decrease two times
the cross section value (see Fig.~\ref{mgg}). For chosen gluon mass
the cross section ratio becomes equal to\[
\sigma (J/\psi +c\bar{c})/\sigma (J/\psi +gg)\approx 0.2\]

It is clear, that the corrections by coused nonzero gluon mass does
not remove disagreement between the pQCD prediction and the experimental
data.

\section{Doubly charmed baryon production}

The research of \( c\bar{c}c\bar{c} \) production would not be complete
without the discussion of the doubly charmed baryon \( \Xi ^{*}_{cc} \).
The \( \Xi ^{*}_{cc}+\bar{c}\bar{c} \) production cross section can
be estimated by the same two methods, which are applied to calculate
the \( J/\Psi +c\bar{c} \) production. 

The first method is based on the factorization theorem. In the frame
work of this approach the \( \Xi ^{*}_{cc} \) production can be represented
as the hard \( c\bar{c}c\bar{c} \) production followed by \( c \)-quark
fusion into the color antitriplet state of \( cc \)-diquark. This
fusion is described by the general coefficient, which is proportional
to the \( cc \)-diquark wave function (it is worth to note that the
situation is analogous to one for \( J/\psi  \) production). At large
energies of \( e^{+}e^{-} \)-interaction the \( cc \)-diquark production
cross section is calculated by convolution of single \( c\bar{c} \)-pair
production with the fragmentation function \( c\rightarrow (cc)_{\bar{3}}+\bar{c} \)
\cite{Falk}. Unfortunately, this asymptotic regime does not set in
at energies, which are researched by BELLE Collaboration. Thus, it
is necessary to take into account the total set of the nonasymptotic
terms. For \( \alpha _{s}=0.24 \) such calculation leads to the following
cross section value : \begin{equation}
\sigma (\Xi ^{*}_{cc})=0.15\pm 0.01\; \mbox {pb}.
\end{equation}
 The ratio of \( \Xi _{cc}^{*} \) production cross section and the
single \( c\bar{c} \)-pair production cross section equals to \begin{equation}
\sigma (\Xi ^{*}_{cc})/\sigma _{c\bar{c}}\sim 10^{-4}.
\end{equation}
 The largest uncertainty of this method is coused by the uncertainty
of \( (cc)_{\bar{3}} \)-diquark wave function. 

\begin{figure}
\( \displaystyle \frac{d\sigma _{(cc)_{\bar{3}}}}{d|\vec{p}_{(cc)_{\bar{3}}}|} \),
pb/GeV 

{\centering \resizebox*{0.7\textwidth}{!}{\includegraphics{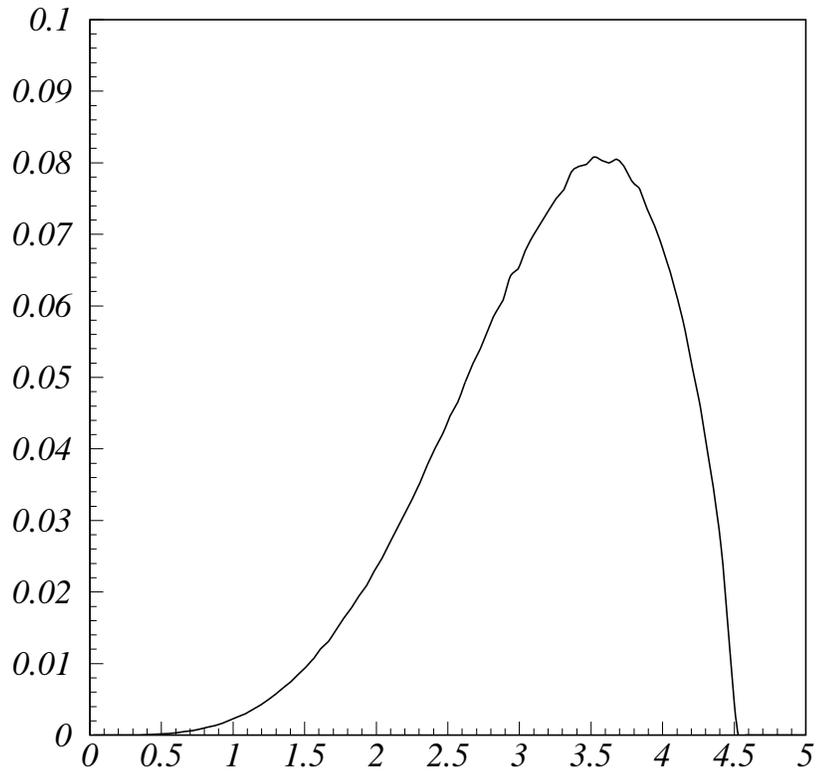}} \par}

\hfill \( |\vec{p}_{(cc)_{\bar{3}}}| \), GeV

\caption{The cross section distribution on the \protect\( (cc)_{\bar{3}}\protect \)-diquark
momentum for the process \protect\( e^{+}e^{-}\rightarrow (cc)_{\bar{3}}+X\protect \)
at \protect\( \sqrt{s}=10.59\protect \) GeV.\label{xicc_p}}
\end{figure}

Another calculation type is based on the quark-hadron duality hypothesis.
In the frame work of this method the diquark production cross section
is evaluated by formula \[
\Sigma \sigma (e^{+}e^{-}\rightarrow (cc)_{\bar{3}}+\bar{c}\bar{c})=\int ^{2m_{c}+\Delta M}_{2m_{c}}\frac{d\sigma }{dM}((cc)_{\bar{3}}+\bar{c}\bar{c})dM,\]
 which results in the following cross section values for the chosen
duality intervals \( \Delta M \): \[
\sigma _{cc}(\Delta M=0.5\; \mbox {GeV})=0.1\; \mbox {pb},\]
\begin{equation}
\sigma _{cc}(\Delta M=1.0\; \mbox {GeV})=0.17\; \mbox {pb}.
\end{equation}
 The expected number of doubly charmed baryons at the luminosity \( L=10^{34} \)cm\( ^{2} \)sec\( ^{-1} \)
is about \( \sim 10^{4} \) pear year. The momentum distribution of
the \( (cc)_{\bar{3}} \)-diquark production cross section is preformed
in Fig.~\ref{xicc_p}.

One can conclude from our discussion, that the cross section value
of \( \Xi ^{*}_{cc}+\bar{c}\bar{c} \) production is close to the
cross section value of \( J/\psi +c\bar{c} \) production. Thus if
one will discover the mechanism, which will explain the large \( J/\psi  \)
production cross section at BELLE experiment, then, it is not unlikely,
that analogous mechanism will predict the large cross section value
of double charmed baryon production. The latter value will be several
times greater than predicted by us in the frame work of pQCD.

We thank V.Kiselev, M.Danilov and P.Pakhlov for the useful discussions. 

This work is supported in part by Grants RFBR 99-02-6558 and 00-15-96645,
Grant of RF Ministry of Education E02-3.1-96 and Grant CRDF MO-001-0.

\end{document}